# Realization of time-reversal invariant photonic topological Anderson insulators

Xiao-Dong Chen[1,†], Zi-Xuan Gao[1,†], Xiaohan Cui[2,†], Hao-Chang Mo[1], Wen-Jie Chen[1], Ruo-Yang Zhang[2], C. T. Chan[2,*] and Jian-Wen Dong[1,*]

1 *School of Physics & State Key Laboratory of Optoelectronic Materials and Technologies, Sun Yat-sen University, Guangzhou 510275, China.*

2 *Department of Physics, The Hong Kong University of Science and Technology, Hong Kong, China.*

[†]These authors contributed equally to this work

[*]Corresponding author: phchan@ust.hk, dongjwen@mail.sysu.edu.cn

**Disorder, which is ubiquitous in nature, has been extensively explored in photonics for understanding the fundamental principles of light diffusion and localization, as well as for applications in functional resonators and random lasers. Recently, the investigation of disorder in topological photonics has led to the realization of topological Anderson insulators characterized by an unexpected disorder-induced phase transition. However, the observed photonic topological Anderson insulators so far are limited to the time-reversal symmetry breaking systems. Here, we propose and realize a photonic quantum spin Hall topological Anderson insulator without breaking time-reversal symmetry. The disorder-induced topological phase transition is comprehensively confirmed through the theoretical effective Dirac Hamiltonian, numerical analysis of bulk transmission, and experimental examination of bulk and edge transmissions. We present the convincing evidence for the unidirectional propagation and robust transport of helical edge modes, which are the key features of nontrivial time-reversal invariant topological Anderson insulators. Furthermore, we demonstrate disorder-induced beam steering, highlighting the potential of disorder as a new degree of freedom to manipulate light propagation in magnetic-free systems. Our work not only paves the way for observing unique topological photonic phases but also suggests potential device applications through the utilization of disorder.**

*Introduction*. --- Disorder, a term denoting the absence of order, permeates various aspects of photonics, serving both as a tool for unraveling fundamental principles of light and as a catalyst for innovative photonic applications [1-4]. For instance, the disorder-induced confinement of light (i.e., Anderson localization) and random lasers in disordered media exemplify its versatility [5, 6]. Nevertheless, in the realm of light transport and its applications in optical communication, disorder is undesirable because it will cause pronounced backscattering and deteriorate the performance of photonic waveguides. As a new avenue to suppressing the backscattering induced by disorders, topological photonics has recently gained significant attention [7-16]. Topological photonic waveguides, supporting gapless chiral or helical edge modes resistant to weak disorders [17-25], have found applications in robust on-chip waveguides and topological lasers [26-35]. However, under sufficiently strong disorders, topological photonic systems become trivial due to the closure of the mobility band gap, leading to strong backscattering as chiral edge transports are blocked by field localization.

Interestingly, instead of being harmful to wave propagation, specific strong disorders can unexpectedly enable robust transport by inducing transitions from topologically trivial phases to a new class of topological phases, called topological Anderson insulators (TAIs) [36-39]. Since their theoretical proposal, photonic topological Anderson insulators (PTAIs) have been experimentally realized in helical waveguides utilizing effective time-varying gauge fields [40], and in gyromagnetic photonic crystals under external magnetic fields [41]. However, the original discovery of TAIs was spurred by investigating disordered quantum spin Hall systems with time-reversal symmetry [36, 37]. The advantage of preserving time-reversal symmetry lies in its potential to achieve disorder-induced physics in magnetic-free systems, making it feasible for realization on the on-chip silicon-on-insulator platform. Thus far, the experimental realization of time-reversal invariant PTAIs remains an unresolved challenge primarily due to two key obstacles: constructing photonic spins within a realistic time-reversal system, and introducing an experimentally feasible parameter to control disorder strength.

Here, we design and realize a time-reversal invariant PTAI by constructing disordered metacrystals that support two oppositely spin-polarized quantum Hall copies that are linked by time-reversal symmetry. Geometrical disorders are introduced by randomly rotating the anisotropic meta-atom in each unit cell. As the disorder strength increases, the metacrystal undergoes a transition from a trivial phase to a nontrivial phase characterized by a nonzero spin Chern number. This disorder-induced topological phase transition is confirmed through not only experimentally measured bulk and edge transmission spectra but also numerical simulations and theoretical analysis. Unidirectional propagation and robust transport of helical edge modes are demonstrated. Furthermore, we demonstrate disorder-induced beam steering, illustrating the practical application in manipulating light propagation through the utilization of engineered disorder.

*Design of time-reversal invariant PTAI.* --- As schematically shown in Fig. 1(a), the time-reversal invariant PTAI supports unidirectional spin-polarized edge modes along its boundary. These edge modes are immune to defects (e.g., sharp corners) and disorders, showing ideal properties for robust electromagnetic wave transport. The unit cell consists of a metallic tripod positioned between two parallel metal plates, each having identical initial spacings of 1.1 mm to the tripod. The structural parameters are lattice constant $a$ = 36.8 mm, height $h$ = 34.6 mm, inner-radius $r$ = 3.65 mm, arm-length $l$ = 7.9 mm, and arm-width $w$ = 2.2 mm [left inset of Fig. 1(a) and Fig. S1]. These parameters are precisely designed to achieve double Dirac cones around K and K' points [42, 43]. This design has two tunable structural parameters controlling the topology of the ordered metacrystal: the orientation angle of the tripod $\alpha$ determining the parity symmetry breaking (PSB) strength, and the vertical displacement of the tripod to the middle plane of the waveguide $d$ determining the spin-orbit coupling (SOC) strength [Fig. 1(b)]. The competition between these two strengths determines whether the metacrystal is topologically trivial or nontrivial. As an example, we consider the ordered metacrystal with $\alpha = 30°$ and $d = 1.1$ mm and examine four bulk bands around 6 GHz [Fig. 1(c)]. A band gap (marked in grey) is found between the second and third bulk bands. To show its topology, we study the eigen fields of four bulk modes below the band gap [Fig. 1(d)]. Due to the combined mirror-z and electromagnetic duality symmetry [44, 45], bulk modes are classified into spin-up mode $|\Psi^\uparrow| = |\sqrt{\varepsilon_0}E_z + \sqrt{\mu_0}H_z|$ and spin-down mode $|\Psi^\downarrow| = |\sqrt{\varepsilon_0}E_z - \sqrt{\mu_0}H_z|$ by checking the phase difference between $E_z$ and $H_z$ fields [20, 21, 23]. Bulk modes are further classified into different orbital modes by examining the rotation direction of their in-plane power flux [46]. As shown in Fig. 1(d), two bulk modes with the same spin at different valleys have opposite orbits, indicating that the metacrystal is characterized by a zero spin Chern number $C_s = (C^\uparrow - C^\downarrow)/2 = 0$ with $C^\uparrow$ ($C^\downarrow$) representing the Chern number of spin-up (spin-down) modes. The topology of an ordered metacrystal can be altered by changing $\alpha$ and $d$, as illustrated in the numerical phase diagram within the region of $\alpha \geq 0$ and $d \geq 0$ [Fig. 1(e)]. The phase diagram within all possible $\alpha$ and $d$ is shown in Fig. S2(a), and bulk bands of four representative ordered metacrystals with their topologies are detailed in Supplement A. The phase diagram exhibits two phases: the trivial one with $C_s = 0$ and the nontrivial one with $C_s = 1$ (i.e., $C^\uparrow = 1$ and $C^\downarrow = -1$). While the aforementioned metacrystal is located within the trivial phase sector (red circle in Fig. 1(e)), it can undergo a topological phase transition by randomly rotating its tripods to suppress the PSB strength. Here, random rotation angles $\theta_i$ of the tripods are uniformly distributed within $[-\theta_d/2, \theta_d/2]$ with the disorder strength indicated by the maximum possible rotation angle $\theta_d$. When $\theta_d$ reaches a sufficient threshold, the strong disorder effectively averages out the PSB effect. Consequently, the PSB strength becomes weaker than the SOC strength, resulting in a disorder-induced phase transition. Notably, increasing $\theta_d$ has the same effect as decreasing $\alpha$ as they both weaken the PSB effect [Fig. S5]. Here, disorder introduces a new degree of freedom for uncovering novel physical phenomena and controlling light propagation, and offers the advantage of avoiding the need for cumbersome and precise control over the structure's morphology. Ultimately, a time-reversal invariant PTAI is achieved as the metacrystal transitions into the nontrivial phase [arrow in Fig. 1(e)].

*Observation of disorder-induced topological phase transition.* --- To prove our theoretical proposal, we first observe the disorder-induced topological phase transition. The fabricated samples consist of a disordered metacrystal with $\theta_d = 0$, 20°, 35°, 60° or 120° and $d = 1.1$ mm placed above an ordered metacrystal with $\alpha = 30°$ and $d = 0$ [Fig. 2(a)]. The lower metacrystal serves as a trivial insulator with $C_s = 0$ inside the band gap from 5.90 to 6.21 GHz. We perform the bulk and edge transmission measurement by putting the source and probe antenna (green and cyan stars) along the boundary and inside the metacrystal, respectively. In Fig. 2(b), the grey shaded rectangles highlight the measured transmission gaps where $|E_z|^2 < 10^{-6}$. With the increase of $\theta_d$, the measured transmission gap initially reduces ($\theta_d = 0$, 20°), then disappears ($\theta_d = 35°$), eventually reopens ($\theta_d = 60°$, 120°). In Fig. 2(c), there is no transmission peak in the band gap for the edge transmission when $\theta_d = 0$, 20° or 35°. In contrast, there is a measured small and big transmission peak when $\theta_d = 60°$ and $\theta_d = 120°$, respectively. These observed characteristics indicate an obvious evidence of a disorder-induced band gap reopening with the topological phase transition. Notably, the subtle differences between the transmission measurements and simulations are attributed to the different probing setups and the presence of minor inherent disorders [Supplement B]. We further conduct numerical bulk transmission to explain the disorder-induced phase transition [Fig. 2(d)]. The numerical setup is schematically shown in Fig. S7(a) and we calculate the rightward power flux. For each disorder strength of $\theta_d$, we consider four different samples with different random configurations of the rotation angles $\theta_i$ and simulate the bulk transmission [Fig. S7(b)]. The averaged bulk transmission in Fig. 2(d) clearly shows the mobility band edges and indicates that the disorder-induced gap closing and reopening occur at approximately $\theta_d = 35°$.

We also build a theoretical model to get a deeper understanding of this topological phase transition. For an ordered metacrystal, four bands at K and K' valleys are described by the Dirac Hamiltonian $\hat{H} = f_D + v_D(\hat{\sigma}_x \hat{\tau}_z \delta k_x + \hat{\sigma}_y \delta k_y) + \eta \hat{\sigma}_z + \kappa \hat{\sigma}_z \hat{s}_z \hat{\tau}_z$, where $\hat{\sigma}_i$, $\hat{\tau}_i$ and $\hat{s}_i$ are the Pauli matrices acting on orbit, valley and spin degrees of freedom, respectively [20, 22, 23, 46]. The fourth (fifth) term describes the band gap opening induced by PSB (SOC) with a gap width coefficient $\eta$ ($\kappa$) [Supplement A]. By averaging the symmetry breaking strengths, we formulate an effective Hamiltonian for the disordered metacrystal:

$$\hat{H}_{\text{disorder}} = f_D + v_D(\hat{\sigma}_x \hat{\tau}_z \delta k_x + \hat{\sigma}_y \delta k_y) + \sum_{i=1}^{i=N} \frac{\eta_i(\theta_i)}{N} \hat{\sigma}_z + \kappa \hat{\sigma}_z \hat{s}_z \hat{\tau}_z \tag{1}$$

where $\eta_i$ is the PSB strength within each unit cell and determined by the rotation angle $\theta_i$, and $\left(\sum_i \eta_i\right)/N$ represents the averaged PSB strength. This effective Hamiltonian implies that the band edges for the disordered metacrystal are given by $f_D \pm \left||\kappa| - \left|\sum_i \eta_i\right|/N\right|$. Two band edges cross when $|\kappa| = \left|\sum_i \eta_i\right|/N$. Before (after) this band crossing, the PSB (SOC) effect is dominant and the metacrystal is trivial with $C_s = 0$ (nontrivial with $C_s = 1$). The theoretical band edges are outlined by two green lines in Fig. 2(d), and match well with the boundary of the low numerical bulk transmission area shown by the background colormap plot. After the mode exchange, the random metacrystal has a band gap which is characterized by $C_s = 1$, hence the time-reversal invariant PTAI is realized. Notably, the spin Bott index [47, 48] is another useful topological invariant to characterize disordered time-reversal systems, but it does not work out here due to the ill-defined spin for modes far away from K and K' points.

*Observation of robust helical edge modes.* --- To substantiate the nontrivial characteristics of the time-reversal invariant PTAI, we demonstrate the unidirectional spin-polarized edge modes of the disordered metacrystal with $\theta_d = 120°$ [see more in Fig. S8]. When the source is put at the left end, only the rightward spin-down edge mode is excited while the rightward spin-up edge mode is forbidden [Fig. 3(a)]. Conversely, when the source is put at the right end, only the spin-up edge mode can propagate leftward [Fig. 3(c)]. This unidirectional spin-polarized edge modes are observed by measuring the transmission spectra of $E_z$ and $H_z$ fields. In Fig. 3(b), high transmission peaks are found in the $|E_z|$ and $|H_z|$ spectra, confirming the excitation of rightward edge mode when the source is on the left. By examining the phase difference (PD) between measured $E_z$ and $H_z$ fields, stable PD around $\pi$ within the band gap indicates that $E_z$ and $H_z$ are out of phase. It means that the rightward edge mode is spin-down polarized, consistent with the numerical result. When the source is put on the right, high transmission is also observed [Fig. 3(d)]. An obvious PD plateau at 0 is measured, signifying the excitation of leftward spin-up edge mode. The clear contrast between two PD plateaus at $\pi$ and 0 indicates the unambiguous observation of the unidirectional spin-polarized edge modes. Here, the light propagation is reciprocal because the time-reversal operator connects modes with opposite spins and opposite momenta [20]. Robust transport of edge modes, another crucial property of the time-reversal invariant PTAI, is also experimentally confirmed [Fig. S9]. In stark contrast to the trivial waveguide, the nontrivial waveguide constructed by the PTAI exhibits broadband robust transport [Fig. S10]. Notably, the finite size effect of the sample is discussed in Supplement F. Expanding the system size and investigating the interplay between disorder effects and Anderson localization is a key objective for future research.

*Observation of disorder-induced beam steering based on topological phase transition*. --- PTAIs, owing to their disorder-controllable nature, can achieve more exotic functionality. Prior investigations about TAIs primarily concentrate on the topological phase transition and the robust transport of edge states. Here, we modulate the flow of light by utilizing disorder as a degree of freedom and demonstrate a previously unobserved phenomenon termed disorder-induced beam steering based on topological phase transition. Specifically, we consider a boundary formed by a trivial metacrystal ($C_s = 0$) at the top and a nontrivial metacrystal ($C_s = -1$) at the bottom [Fig. 4(a)]. This boundary supports rightward spin-down (leftward spin-up) edge mode at the K' (K) valley [Fig. 4(b)]. When the source is put at the left, only the rightward spin-down edge mode at the K' valley is excited. According to the k-space analysis [inset of Fig. 4(a)], the excited edge mode refracts downwards into the air waveguide, resulting in a single light beam emission at the air-metacrystal interface [details in Fig. S12]. This prediction is confirmed by the full-wave simulation where only one refracted plane wave is observed, and is further corroborated by experimental scanning of $H_z$ fields in the air waveguide [Fig. 4(c)]. We then introduce disorder into the upper metacrystal and keep the lower metacrystal unchanged. When the disorder strength is $\theta_d = 120°$, the upper metacrystal becomes a PTAI with $C_s = 1$ [Fig. 4(d)]. As the difference of spin Chern number crossing the boundary is -2, this boundary supports two spin-up (spin-down) edge modes with negative (positive) group velocity [Fig. 4(e)]. Therefore, the left source excites rightward spin-down edge modes at both K and K' valleys, resulting in two refracted light beams [inset of Fig. 4(d)]. The simulated and measured $H_z$ fields show that there are two out-coupling light beams [Fig. 4(f)]. Notably, these two light beams are induced by introducing disorder rather than switching the polarization. With the clear contrast between refracted light beams in Figs. 4(c) and 4(f), we demonstrate that disorder can manipulate light propagation.

*Conclusion*. --- We realize a time-reversal invariant PTAI where strong geometric disorder smooths out the PSB effect and drives an initial trivial system into a nontrivial photonic quantum spin Hall phase. The disorder-induced topological phase transition is evidenced by the experimental and theoretical results, including the closing and reopening of transmission gaps, the emergence of edge modes, and the phase diagram based on the effective Dirac Hamiltonian. The unidirectional propagation and robust transport of helical edge modes are experimentally observed. Furthermore, we demonstrate the light beam steering leveraged by disorder-induced topological phase transition. This work provides valuable insights for observing unique topological photonic phases in time-reversal invariant (i.e, magnetic-free, nonlinearity-free, and time-modulation-free) systems. It also suggests potential device applications (e.g., beam splitters) through the utilization of disorder.

**Acknowledgements**

This work was supported by National Key Research Development Program of China (2022YFA1404304), National Natural Science Foundation of China (12074443, 12374364, 62035016), Guangdong Basic and Applied Basic Research Foundation (2023B1515040023), Guangzhou Science, Technology and Innovation Commission (2024A04J6333), Fundamental Research Funds for the Central Universities of the Sun Yat-sen University (23lgbj021), and by Hong Kong Research Grants Council through Grant Nos. 16307621, 16310422, AoE/P-502/20.

# REFERENCES

[1] D. S. Wiersma, "Disordered photonics," Nature Photonics **7**, 188-196 (2013).

[2] S. Yu, C.-W. Qiu, Y. Chong, S. Torquato, and N. Park, "Engineered disorder in photonics," Nature Reviews Materials **6**, 226-243 (2021).

[3] X. Jiang, L. Shao, S.-X. Zhang, X. Yi, J. Wiersig, L. Wang, Q. Gong, M. Lončar, L. Yang, and Y.-F. Xiao, "Chaos-assisted broadband momentum transformation in optical microresonators," Science **358**, 344-347 (2017).

[4] X. Jiang, S. Yin, H. Li, J. Quan, H. Goh, M. Cotrufo, J. Kullig, J. Wiersig, and A. Alù, "Coherent control of chaotic optical microcavity with reflectionless scattering modes," Nature Physics **20**, 109-115 (2024).

[5] P. W. Anderson, "Absence of Diffusion in Certain Random Lattices," Physical Review **109**, 1492-1505 (1958).

[6] M. Lee, S. Callard, C. Seassal, and H. Jeon, "Taming of random lasers," Nature Photonics **13**, 445-448 (2019).

[7] L. Lu, J. D. Joannopoulos, and M. Soljačić, "Topological photonics," Nature Photonics **8**, 821 (2014).

[8] A. B. Khanikaev, and G. Shvets, "Two-dimensional topological photonics," Nature Photonics **11**, 763-773 (2017).

[9] T. Ozawa, H. M. Price, A. Amo, N. Goldman, M. Hafezi, L. Lu, M. C. Rechtsman, D. Schuster, J. Simon, O. Zilberberg, and I. Carusotto, "Topological photonics," Reviews of Modern Physics **91**, 015006 (2019).

[10] M. Kim, Z. Jacob, and J. Rho, "Recent advances in 2D, 3D and higher-order topological photonics," Light: Science & Applications **9**, 130 (2020).

[11] E. Lustig, and M. Segev, "Topological photonics in synthetic dimensions," Advances in Optics and Photonics **13**, 426 (2021).

[12] S. Ma, B. Yang, and S. Zhang, "Topological photonics in metamaterials," Photonics Insights **1**, R02 (2022).

[13] M. Jalali Mehrabad, S. Mittal, and M. Hafezi, "Topological photonics: Fundamental concepts, recent developments, and future directions," Physical Review A **108**, 040101 (2023).

[14] J. W. You, Z. Lan, Q. Ma, Z. Gao, Y. Yang, F. Gao, M. Xiao, and T. J. Cui, "Topological metasurface: from passive toward active and beyond," Photonics Research **11**, B65 (2023).

[15] X. Zhang, F. Zangeneh-Nejad, Z. G. Chen, M. H. Lu, and J. Christensen, "A second wave of topological phenomena in photonics and acoustics," Nature **618**, 687-697 (2023).

[16] N. Han, X. Xi, Y. Meng, H. Chen, Z. Gao, and Y. Yang, "Topological photonics in three and higher dimensions," APL Photonics **9**, 010902 (2024).

[17] Z. Wang, Y. Chong, J. D. Joannopoulos, and M. Soljačić, "Observation of unidirectional backscattering-immune topological electromagnetic states," Nature **461**, 772-775 (2009).

[18] Y. Poo, R.-x. Wu, Z. Lin, Y. Yang, and C. T. Chan, "Experimental realization of self-guiding unidirectional electromagnetic edge states," Physical Review Letters **106**, 093903 (2011).

[19] C. He, X.-C. Sun, X.-P. Liu, M.-H. Lu, Y. Chen, L. Feng, and Y.-F. Chen, "Photonic topological insulator with broken time-reversal symmetry," Proceedings of the National Academy of Sciences **113**, 4924-4928 (2016).

[20] A. B. Khanikaev, S. H. Mousavi, W.-K. Tse, M. Kargarian, A. H. MacDonald, and G. Shvets,

"Photonic topological insulators," Nature Materials **12**, 233 (2012).

[21] W.-J. Chen, S.-J. Jiang, X.-D. Chen, B. Zhu, L. Zhou, J.-W. Dong, and C. T. Chan, "Experimental realization of photonic topological insulator in a uniaxial metacrystal waveguide," Nature communications **5**, 5782 (2014).

[22] T. Ma, A. B. Khanikaev, S. H. Mousavi, and G. Shvets, "Guiding electromagnetic waves around sharp corners: topologically protected photonic transport in metawaveguides," Physical Review Letters **114**, 127401 (2015).

[23] X. Cheng, C. Jouvaud, X. Ni, S. H. Mousavi, A. Z. Genack, and A. B. Khanikaev, "Robust reconfigurable electromagnetic pathways within a photonic topological insulator," Nature Materials **15**, 542-548 (2016).

[24] L.-H. Wu, and X. Hu, "Scheme for achieving a topological photonic crystal by using dielectric material," Physical Review Letters **114**, 223901 (2015).

[25] H. Yang, J. Xu, Z. Xiong, X. Lu, R.-Y. Zhang, H. Li, Y. Chen, and S. Zhang, "Optically Reconfigurable Spin-Valley Hall Effect of Light in Coupled Nonlinear Ring Resonator Lattice," Physical Review Letters **127**, 043904 (2021).

[26] S. Barik, A. Karasahin, C. Flower, T. Cai, H. Miyake, W. DeGottardi, M. Hafezi, and E. Waks, "A topological quantum optics interface," Science **359**, 666-668 (2018).

[27] M. I. Shalaev, W. Walasik, A. Tsukernik, Y. Xu, and N. M. Litchinitser, "Robust topologically protected transport in photonic crystals at telecommunication wavelengths," Nature Nanotechnology **14**, 31-34 (2018).

[28] X.-T. He, E.-T. Liang, J.-J. Yuan, H.-Y. Qiu, X.-D. Chen, F.-L. Zhao, and J.-W. Dong, "A silicon-on-insulator slab for topological valley transport," Nature communications **10**, 872 (2019).

[29] D. a. J. Bisharat, and D. F. Sievenpiper, "Valley Polarized Edge States beyond Inversion Symmetry Breaking," Laser & Photonics Reviews **17**, 2200362 (2023).

[30] Y. Kawaguchi, D. Smirnova, F. Komissarenko, S. Kiriushechkina, A. Vakulenko, M. Li, A. Alù, and A. B. Khanikaev, "Pseudo-spin switches and Aharonov-Bohm effect for topological boundary modes," Science advances **10**, eadn6095 (2024).

[31] B. B. A. Ndao, F. Vallini, A. E. Amili, Y. Fainman, and B. Kanté, "Nonreciprocal lasing in topological cavities of arbitrary geometries," Science **358**, 636-640 (2017).

[32] M. A. Bandres, S. Wittek, G. Harari, M. Parto, J. Ren, M. Segev, D. N. Christodoulides, and M. Khajavikhan, "Topological insulator laser: Experiments," Science **359**, eaar4005 (2018).

[33] Z. K. Shao, H. Z. Chen, S. Wang, X. R. Mao, Z. Q. Yang, S. L. Wang, X. X. Wang, X. Hu, and R. M. Ma, "A high-performance topological bulk laser based on band-inversion-induced reflection," Nat Nanotechnol (2019).

[34] Y. Zeng, U. Chattopadhyay, B. Zhu, B. Qiang, J. Li, Y. Jin, L. Li, A. G. Davies, E. H. Linfield, B. Zhang, Y. Chong, and Q. J. Wang, "Electrically pumped topological laser with valley edge modes," Nature **578**, 246-250 (2020).

[35] J. Ma, T. Zhou, M. Tang, H. Li, Z. Zhang, X. Xi, M. Martin, T. Baron, H. Liu, Z. Zhang, S. Chen, and X. Sun, "Room-temperature continuous-wave topological Dirac-vortex microcavity lasers on silicon," Light Sci Appl **12**, 255 (2023).

[36] J. Li, R.-L. Chu, J. K. Jain, and S.-Q. Shen, "Topological Anderson Insulator," Physical Review Letters **102**, 136806 (2009).

[37] C. W. Groth, M. Wimmer, A. R. Akhmerov, J. Tworzydło, and C. W. J. Beenakker, "Theory of the Topological Anderson Insulator," Physical Review Letters **103**, 196805 (2009).

[38] C. Liu, W. Gao, B. Yang, and S. Zhang, "Disorder-Induced Topological State Transition in Photonic Metamaterials," Physical Review Letters **119**, 183901 (2017).
[39] W. Zhang, D. Zou, Q. Pei, W. He, J. Bao, H. Sun, and X. Zhang, "Experimental Observation of Higher-Order Topological Anderson Insulators," Physical Review Letters **126**, 146802 (2021).
[40] S. Stutzer, Y. Plotnik, Y. Lumer, P. Titum, N. H. Lindner, M. Segev, M. C. Rechtsman, and A. Szameit, "Photonic topological Anderson insulators," Nature **560**, 461-465 (2018).
[41] G.-G. Liu, Y. Yang, X. Ren, H. Xue, X. Lin, Y.-H. Hu, H.-x. Sun, B. Peng, P. Zhou, Y. Chong, and B. Zhang, "Topological Anderson Insulator in Disordered Photonic Crystals," Physical Review Letters **125**, 133603 (2020).
[42] F. Gao, H. Xue, Z. Yang, K. Lai, Y. Yu, X. Lin, Y. Chong, G. Shvets, and B. Zhang, "Topologically protected refraction of robust kink states in valley photonic crystals," Nature Physics **14**, 140-144 (2018).
[43] H. Xue, F. Gao, Y. Yu, Y. Chong, G. Shvets, and B. Zhang, "Spin-valley-controlled photonic topological insulator," arXiv **1811.00393** (2018).
[44] M. G. Silveirinha, "P·T·D symmetry-protected scattering anomaly in optics," Physical Review B **95**, 035153 (2017).
[45] X. Cui, R.-Y. Zhang, Z.-Q. Zhang, and C. T. Chan, "Photonic Z2 Topological Anderson Insulators," Physical Review Letters **129**, 043902 (2022).
[46] T. Ma, and G. Shvets, "Scattering-free edge states between heterogeneous photonic topological insulators," Physical Review B **95**, 165102 (2017).
[47] H. Huang, and F. Liu, "Quantum Spin Hall Effect and Spin Bott Index in a Quasicrystal Lattice," Phys Rev Lett **121**, 126401 (2018).
[48] H. Huang, and F. Liu, "Theory of spin Bott index for quantum spin Hall states in nonperiodic systems," Physical Review B **98** (2018).

**Figures and captions**

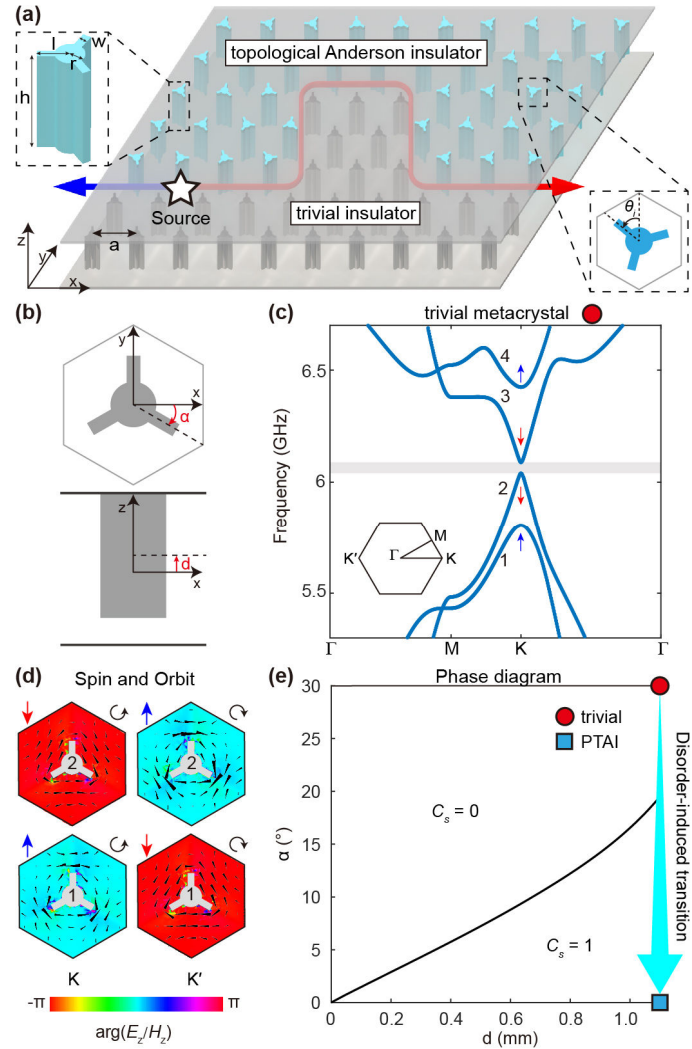

FIG. 1. (a) Schematic of the disordered metacrystal with the tripods rotated by random angles, forming a PTAI. At the boundary between the PTAI and a trivial insulator, robust spin-up and spin-down edge modes propagate in opposite directions. A zoom-in metallic tripod and a rotated tripod with $\theta_i$ are shown as insets. (b) The angle ($\alpha$) induces the parity symmetry breaking (PSB) effect, and the distance ($d$) introduces the spin-orbit coupling (SOC) effect. (c) Bulk bands of an ordered metacrystal with $\alpha = 30°$ and $d = 1.1$ mm. Four bulk modes are marked with their spins. (d) Phase difference (PD) between $E_z$ and $H_z$ (colors) and power flux (black arrows) of bulk modes below the band gap. (e) Numerical phase diagram for ordered metacrystals within the space of two parameters $\alpha$ and $d$. The trajectory for realizing a PTAI is indicated by a cyan arrow. Sufficiently strong disorder drives the metacrystal into a PTAI.

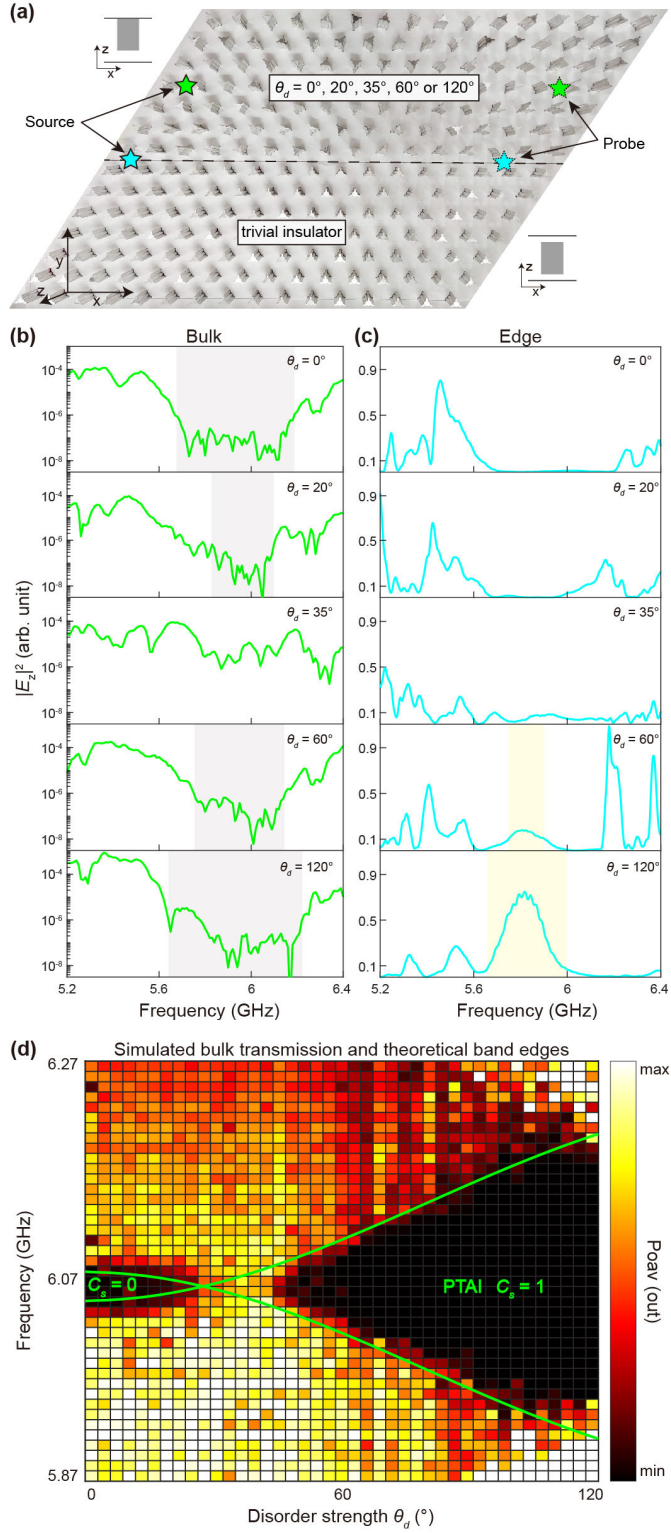

FIG. 2. (a) Sample between the disordered metacrystal with $\theta_d$ and $d = 1.1$ mm (top) and a trivial insulator with $\alpha = 30°$ and $d = 0$ (bottom). The green and cyan stars mark antenna for measuring the bulk and edge transmission spectra. (b-c) Measured bulk and edge transmission spectra. Grey rectangles in (b) mark the measured bulk band gaps. Yellow rectangles in (c) mark the high transmission frequency range where edge modes appear. The edge transmission spectra are presented using linear coordinates for improved clarity. (d) Simulated bulk transmittance averaged over four samples (background density plot) and the theoretical band edges (green curves).

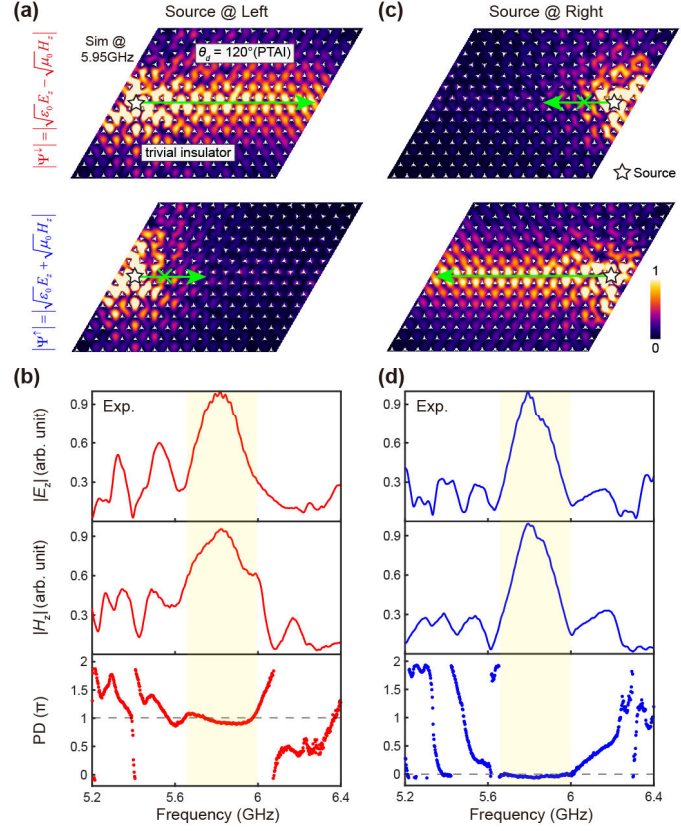

FIG. 3. (a) Simulated local field intensity $|\psi^\downarrow|$ (top) and $|\psi^\uparrow|$ (bottom) and (b) Measured $|E_z|$, $|H_z|$, phase difference (PD) between $E_z$ and $H_z$ fields when the source is put at the left. (c-d) Simulated and measured results when the source is put at the right. An obvious PD plateau at $\pi$ (0) is observed when the source is put at left (right), indicating the unidirectional propagation of spin-polarized edge modes.

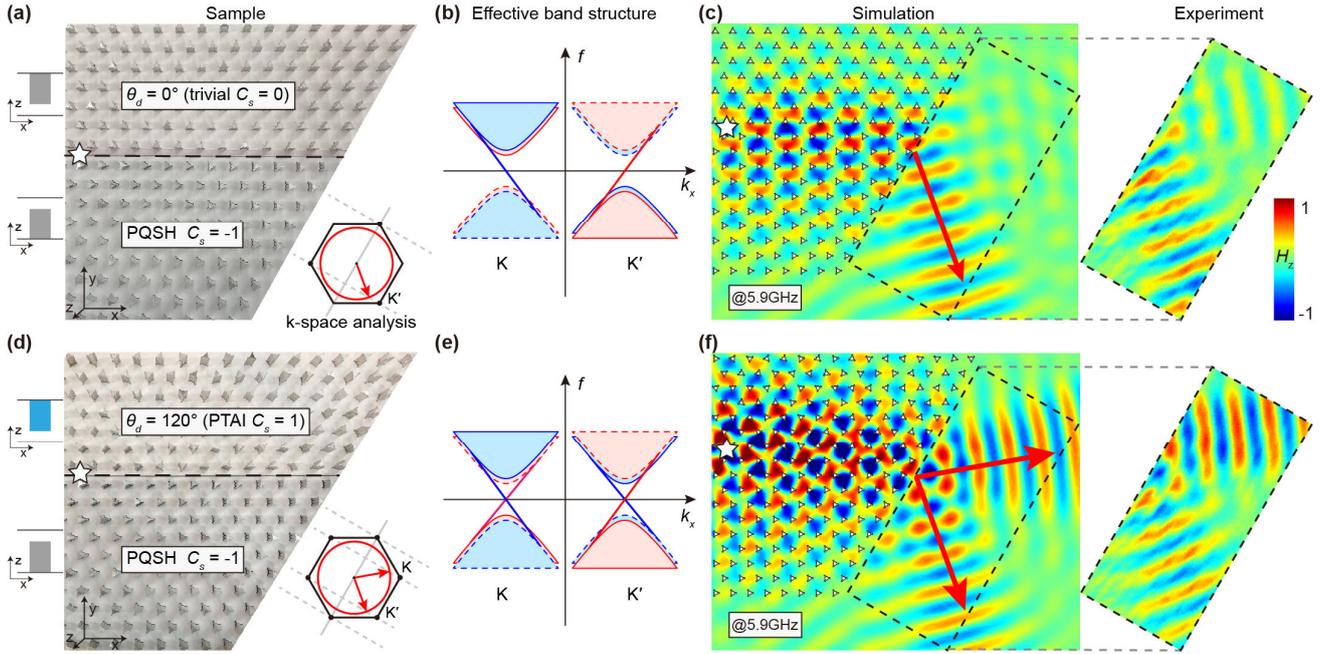

FIG. 4. (a-b) Photo and effective projected band structure of the sample consisting of the metacrystal with $\theta_d = 0°$ ($C_s = 0$) and the photonic quantum spin Hall (PQSH) metacrystal with $\alpha = 0°$ and $d = -1.1$ mm ($C_s = -1$). (c) The simulated and measured $H_z$ fields for the refraction of edge modes through the zigzag termination. (d-f) Case for the PTAI with $\theta_d = 120°$ ($C_s = 1$). Inset of (a, d): The k-space analysis on the refraction of edge modes.